\documentclass[]{emulateapj}
\begin{document}

\title{Determining the evolutionary stage of the $\delta$ Scuti star HIP 80088 with asteroseismology}
\author{Xinghao Chen\altaffilmark{1,2} and Yan Li \altaffilmark{1,2,3,4}}
%\email{chenxinghao@ynao.ac.cn; ly@ynao.ac.cn}

\altaffiltext{1}{Yunnan Observatories, Chinese Academy of Sciences, P.O. Box 110, Kunming 650216, China; chenxinghao@ynao.ac.cn; ly@ynao.ac.cn}
\altaffiltext{2}{Key Laboratory for Structure and Evolution of Celestial Objects, Chinese Academy of Sciences, P.O. Box 110, Kunming 650216, China}
\altaffiltext{3}{University of Chinese Academy of Sciences, Beijing 100049, China}
\altaffiltext{4}{Center for Astronomical Mega-Science, Chinese Academy of Sciences, 20A Datun Road, Chaoyang District, Beijing, 100012, China}
\begin{abstract}
We have computed a grid of theoretical models to fit the twelve oscillation modes of HIP 80088 observed by K2. HIP 80088 is determined to be a pre-main sequence star, in which CN cycle has not arrived at the equilibrium state. Mass fractions of C12 and N14 in metal composition are 0.1277$^{+0.0064}_{-0.0049}$ and 0.1092$^{+0.0057}_{-0.0074}$ respectively, indicating that twenty-eight percent of C12 have turned into N14. Meanwhile, our fitting results show that physical parameters of HIP 80088 converge to a small range: $M$ = 1.68$-$1.78 $M _{\odot}$, $Z$ = 0.015$-$0.018, $\upsilon_{\rm e}$ = 120$-$136 km s$^{-1}$, $\log g$ = 4.114$-$4.125, $R$ = 1.882$-$1.914 $R_{\odot}$, $\tau_{0}$ = 7636$-$7723 s, and age = 9.03$-$10.21 Myr. Based on our model fittings, $f_{3}$ is suggested to be one radial mode, $f_{2}$, $f_{4}$, $f_{8}$, and $f_{11}$ to be four $\ell$ = 1 modes, and $f_{1}$, $f_{5}$, $f_{6}$, $f_{7}$, $f_{9}$, $f_{10}$, and $f_{12}$ to be seven $\ell$ = 2 modes. In particularly, we find that ($f_{2}$, $f_{4}$, $f_{8}$) form one complete triplet with the averaged frequency spacing of 16.045 $\mu$Hz , and ($f_{5}$, $f_{7}$, $f_{9}$, $f_{10}$) form four components of one quintuplet with the averaged frequency spacing of 13.388 $\mu$Hz. The two averaged frequency spacings are not equal. Based on the best-fitting model, those $\ell$ = 2 modes of HIP 80088 are found to be mixed modes, which are p-dominated modes with pronounced g-mode features, while oscillation modes with $\ell$ = 1 are p modes.
\end{abstract}


\section{Introduction}
The $\delta$ Scuti stars are a class of A- and F-type stars with masses in the range of 1.5$-$2.5 $M_{\odot}$, located in the intersection region of the main sequence and the classical instability strip. They are mainly in main sequence (MS) and post-main sequence (POMS) (Breger et al. 2000; Aerts et al. 2010), but also in pre-main sequence stage (PMS) (Breger 1972; Zwintz 2008; Zwintz et al. 2011; Zwintz et al. 2014). Since evolutionary tracks of PMS, MS, and POMS usually intersect each other on Hertzsprung-Russel diagram, it is too difficult to distinguish the evolutionary stage of a star only on basis of its mass, effective temperature, and luminosity. So far, there are two common ways to resolve this ambiguity. One way is on basis of typical observational characters, such as belonging to a young association or cluster. The other way is the method of asteroseismology, by comparing the theoretical predicted frequencies and the observed oscillation modes (Guenther et al. 2007; Zwintz et al. 2014). In particular, asterosesimic study not only can provide an independent determination of fundamental physical parameters of pulsating stars, but also can investigate more detailed information of their interior.

HIP 80088 (EPIC 204372172) was identified as a $\delta$ Scuti star by Ripepi et al. (2015) based on an analysis of the periodogram. Moreover, Ripepi et al. (2015) obtained 12 independent oscillation frequencies from the kepler photometric timeseries, which were observed from 2014 August 23 to 2014 November 10 for 78.7 days in short-cadence mode (one-min exposures) during the K2-C02 campaign. The oscillation frequencies are listed in Table 1.

In addition, Pecaut et al. (2012) show HIP 80088 an A9V star with the effective temperature $T_{\rm eff}$ = 7500 $\pm$ 150 K and luminosity log($L/L_{\odot}$) = 0.960 $\pm$ 0.095. In 2015, Ripepi et al. obtained one-hour high-resolution spectra with the Catania Astrophysical Observatory Spectropolarimeter mounted on the Cassegrain fcous of the 91-cm telescope of the 'M. G. Fracastoro' observing station of the Catania Astrophysical Observatory.  From the spectra, Ripepi et al. (2015) obtained that values of $T_{\rm eff}$, $\log g$, and $\upsilon$sin$i$ are 7600 $\pm$ 200 K, 3.80 $\pm$ 0.15, and 80 $\pm$ 5 km s$^{-1}$ respectively, using the spectrum synthesis approach (Kurucz 1993a,b; Kurucz $\&$ Avrett 1981; Sbordone et al. 2004). 

Furthermore, de Zeeuw et al. (1999) identified HIP 80088 as a member star of the Upper Scorpius association (USco) on basis of  a moving group analysis of the Hipparcos positions, parallaxes, and proper motions. The Upper Scorpius association is the youngest subgroup of the Scorpius-Centaurus association, and its median age is estimated to 
be 5 Myr for K and M-type stars (Preibisch et al. 2002; Slesnick et al. 2008; Feiden 2016) and 10$-$13 Myr for stars with spectral types of B, A, F, and G (Pecaut et al 2012; Ripepi et al. 2015; Feiden 2016). Ripepi et al. (2015) showed HIP 80088 most likely to be a pre-main sequence star or else a star very close to zero-age main sequence. Besides, HIP 80088 is expected to share a common age and chemical composition as a member star of USco. Both of spectral and photometric observations are available, which make HIP 80088 an ideal target for asteroseismic analyses.  In Section 2, we report our stellar models. Stellar evolution code and input physics are described in Section 2.1, grids of theoretical models are presented in Section 2.2, and details of model fittings are introduced in Section 2.3. We present the fitting results in Section 3, and discuss them in Section 4. Finally, the main results of our work are concluded in Section 5.
\section{Stellar models}
\subsection{Stellar evolution code and input physics}
In our work, we use the one-dimensional stellar evolution code Modules for Experiments in Stellar Astrophysics (MESA; Paxton et al. 2011, 2013) to generate theoretical models. The sub-module pulse from version 6596 is used to compute stellar evolutionary models and their corresponding  oscillation modes (Christensen-Dalsgaard 2008; Paxton et al. 2011, 2013)

In our calculations, the 2005 update of the OPAL  equation of state  (Rogers $\&$ Nayfonov 2002) is adopted. The OPAL opacity tables from Iglesias $\&$ Rogers (1996) are used in the high temperature region, whereas in the lower temperature region tables from Ferguson et al. (2005) are used instead. The initial composition in metal is assumed to be the same as the solar metal composition AGSS09 (Asplund et al. 2009). The $T - \tau$ relation of Eddington grey atmosphere is chosen in the atmosphere integration. The mixing-length theory (MLT) of B$\ddot{\rm o}$hm-Vitense (1958) with $\alpha$ = 1.8 is adopted to treat convection. In addition, effects of convective overshooting, element diffusion, and rotation on the structure and evolution of the star are not considered in our work.

\subsection{Grid of theoretical models}
We use the MESA code to generate a grid of stellar models. The mass $M$ varies from 1.5 $M_{\odot}$ to 2.5 $M_{\odot}$ with a step of 0.02 $M_{\odot}$.  The mass fraction of heavy-elements $Z$ varies from 0.010 to 0.030 with a step of 0.001. The initial helium fraction is set to be $Y=0.245+1.54Z$ (e.g., Dotter et al. 2008; Thompson et al. 2014) as a function of $Z$.

Each star in the grid evolves starting from the pre-main sequence and ending when its age reaches to 200 Myr.  The rectangle in Figure 1 marks 1 $\sigma$ error box of the observed stellar parameters, the effective temperature 7350 K $<$ $T_{\rm eff}$ $<$ 7800 K (Pecaut et al. 2012; Ripepi et al. 2015) and the luminosity  0.865 $<$ log($L/L_{\odot}$) $<$ 1.055 (Pecaut et al. 2012). As a star evolves along its evolutionary track like Figure 1, we calculate frequencies of oscillation modes with $\ell$ = 0, 1, and 2 for every stellar model falling inside the rectangle.
 \subsection{Model fittings}
According to the theory of stellar oscillations, each oscillation mode can be described by three indices ($n$, $\ell$, $m$), which denote the radial order, the spherical harmonic degree, and the azimuthal number, respectively. If a star is spherical symmetry, the azimuthal number $m$ is degenerate, which means that modes with the same $n$ and $\ell$ but different $m$ have the same frequency.  However for a rotating star, effects of rotation break the star$'$s spherical symmetry and result in one nonradial oscillation mode splitting into $2\ell+1$ different ones. Aerts et al. (2010) derived the general expression of the first-order effect of rotation as
\begin{equation}
\nu_{\ell,n,m}= \nu_{\ell,n} + \beta_{\ell, n}\frac{m}{P_{\rm rot}} =  \nu_{\ell,n} + \beta_{\ell, n}\frac{m \upsilon_{\rm e}}{2\pi R} ,
\end{equation}
where $P_{\rm rot}$ is the rotational period, $R$ the stellar radius, $\upsilon_{\rm e}$ the equatorial rotational velocity, $\beta_{n,\ell}$ the rotational parameter measuring the size of rotational splitting, and $m$ ranging from $-\ell$ to $\ell$ in a step of 1.  For a uniformly rotating star, Aerts et al. (2010) showed the general expression of $\beta_{n,\ell}$ as
\begin{equation}
\beta_{\ell, n}=\frac{\int_{0}^{R}(\xi_{r}^{2}+L^{2}\xi_{h}^{2}-2\xi_{r}\xi_{h}-\xi_{h}^{2})r^{2}\rho dr}
{\int_{0}^{R}(\xi_{r}^{2}+L^{2}\xi_{h}^{2})r^{2}\rho dr}.
\end{equation}
In equation (2), $\xi_{r}$ is the radial displacement, $\xi_{h}$ the horizontal displacement, $\rho$ the local density, and $L^{2}= \ell(\ell+1)$.

The projected rotational velocity $\upsilon$sin$i$ of HIP 80088 is obtained to be 80 $\pm$ 5 km s$^{-1}$ (Ripepi et al. 2015). Whereas, the inclination angle $i$ is unknown. In our work, we consider the equatorial rotation velocity $\upsilon_{\rm e}$ as one adjustment parameter.  The value of $\upsilon_{\rm e}$ ranges from 50 to 200 with a step of 5 in unit of kilometer per second. For a given $\upsilon_{\rm e}$, each nonradial oscillation mode calculated in Section 2.2 will split into several ones on basis of equation (1). Namely, every mode with $\ell$ = 1 splits into three different frequencies, which correspond to $m$ = -1, 0, and 1, respectively.  Every mode with $\ell$ = 2 splits into five different frequencies, which correspond to $m$ = -2, -1, 0, 1, and 2, respectively.

Then we perform a $\chi^{2}$ minimization by fitting them to observations according to
\begin{equation}
\chi^{2}=\frac{1}{k}\sum(|\nu_{i}^{\rm obs}-\nu_{i}^{\rm theo}|^{2}).
\end{equation}
The smaller the value of $\chi^{2}$, the higher the probability of matching the observations. In Equation (3), $\nu_{i}^{\rm obs}$ and $\nu_{i}^{\rm theo}$ are the observed frequency and the theoretical calculated frequency respectively, and k is the number of observed oscillation modes. Moreover, it should be pointed out that the observed oscillation modes are not identified in advanced. When doing model fittings, the theoretical frequency nearest to  the observed frequency is treated as its possible model counterpart.

\section{Fitting results of HIP 80088}
Figures 2 and 3 present plots of $\chi_{\rm m}^{2}$ to various physical parameters. Thereinto, each circle denotes one minimum value $\chi_{\rm m}^{2}$ of $\chi^{2}$ on one evolutionary track. The circles falling into the rectangle correspond to twenty-nine candidate models listed in Table 2, in particular, the filled circle marks the best-fitting model (Model 29).

Figures 2(a) - (c) illustrate $\chi_{\rm m}^{2}$ as a function of the stellar mass $M$, the mass fraction of heavy-elements $Z$, and the equatorial rotation velocity $\upsilon_{\rm e}$, respectively. It can be seen in Figures (a)-(c) that the three adjustable parameters ($M$, $Z$, $\upsilon_{\rm e}$) are well limited in a small parameter space, i.e., $M$ = 1.68 $-$ 1.78 $M_{\odot}$,  $Z$ = 0.015 $-$ 0.018, and $\upsilon_{\rm e}$ = 120 $-$ 136 km s$^{-1}$. The projected rotational velocity $\upsilon$sin$i$ is determined to be 80 $\pm$ 5 km s$^{-1}$ (Ripepi et al. 2015). Then the inclination angle $i$ is suggested to be 38.3 $_{-4.8}^{+6.8}$ degree.

Figure 2(d) illustrates $\chi_{\rm m}^{2}$ as a function of ages of stars. In Figure 2(d), we find that ages of the candidate models converge to 9.03 $-$ 10.21 Myr, which is in accordance with the result of Ripepi et al. (2015).

Figures 3(a) and (b) depict $\chi_{\rm m}^{2}$ as a function of the gravitational acceleration $\log g$ and the radius $R$, respectively. Figures 3(a) shows that $\log g$ converges to 4.114 $-$ 4.125.  Figure 3(b) shows that $R$ converges to 1.882 $-$ 1.914 $R_{\odot}$.

Figures 3(c) depicts $\chi_{\rm m}^{2}$ as a function of the acoustic radius $\tau_{0}$. The acoustic radius $\tau_{0}$ is the sound travel time between the center and the surface of the star. It is defined by Aerts et al. (2010) as
\begin{equation}
\tau_{0}=\int_{0}^{R}\frac{dr}{c_{\rm s}},
\end{equation}
where $R$ is the stellar radius, and $c_{\rm s}$ is the adiabatic sound speed. The acoustic radius $\tau_{0}$ is one significant asteroseismic quantity, which is used to characterize the features in the envelope of the star(e.g., Ballot et al. 2004; Miglio et al. 2010; Chen et al. 2016, 2017; Wu $\&$ Li 2016; Chen $\&$ Li 2017). In Figure 3(c), it can be found that $\tau_{0}$ converges well to 7636 $-$ 7732 s.

Figures 3(d) and (e) depict $\chi_{\rm m}^{2}$ as a function of C12/$Z$ and N14/$Z$, respectively. Thereinto, C12/$Z$ is the mass fraction of C12 in metal composition, and  N14/$Z$ is the mass fraction of N14 in metal composition. In our work, the initial composition in metal is assumed to be the same as the solar metal composition AGSS09 (Asplund et al. 2009). The initial values of C12/$Z$ and N14/$Z$ are 0.1769 and 0.0518 respectively. From Figures 3(d) and (e), HIP 80088 is determined to be a pre-main sequence star, in which CN cycle has not reached the equilibrium state. The mass fraction of C12 in metal composition is determined to be 0.1277$^{+0.0064}_{-0.0049}$ and that of N14 be 0.1092$^{+0.0057}_{-0.0074}$. Twenty-eight percent of C12 have turned into N14.

Based on above considerations, we obtain fundamental physical parameters of HIP 80088 and list them in Table 3, where the best-fitting model is marked in boldface. As for the best-fitting model, the theoretical calculated frequencies are listed in Table 4, and the comparing results between theoretical calculated frequencies and the twelve observed frequencies are listed in Table 5. It can be found  in Table 5 that $f_{3}$ is identified as one radial mode, $f_{2}$, $f_{4}$, $f_{8}$, and $f_{11}$ as four $\ell$ = 1 modes, and $f_{1}$, $f_{5}$, $f_{6}$, $f_{7}$,$f_{9}$, $f_{10}$, and $f_{12}$ as seven $\ell$ = 2 modes. In particularly, we find that ($f_{2}$, $f_{4}$, $f_{8}$) form one complete triplet, and ($f_{5}$, $f_{7}$, $f_{9}$, $f_{10}$) form four components of one quintuplet. We list the two sets of possible rotational splittings in Table 6.

\section{Discussions}
In Section 3, we elaborated our fitting results. Two sets of possible rotational splittings are identified with the method of model fittings. Meanwhile, we noticed in Table 6 that the averaged frequency spacing in ($f_{2}$, $f_{4}$, $f_{8}$) is 16.045 $\mu$Hz, and that in ($f_{5}$, $f_{7}$, $f_{9}$, $f_{10}$) is 13.388 $\mu$Hz. The two averaged frequency spacings are not equal. In order to explain this, propagating properties of the oscillation modes in the star are examined in detail.

Figures 4 and 5 illustrate the behaviors of Brunt$-$V$\ddot{\rm a}$is$\ddot{\rm a}$l$\ddot{\rm a}$ frequency $N$ and characteristic acoustic frequencies $S_{\ell}$ ($\ell$ = 1 and 2) inside the best-fitting model of HIP 80088. The regions with $\omega^{2}$ $>$ $N^{2}$ and $\omega^{2}$ $>$ $S_{\ell}^{2}$ are the propagation zones of p modes. The regions with $\omega^{2}$ $<$ $N^{2}$ and $\omega^{2}$ $<$ $S_{\ell}^{2}$ are the propagation zones of g modes. The regions with $S_{\ell}^{2}$ $<$ $\omega^{2}$ $<$ $N^{2}$ or $N^{2}$ $<$ $\omega^{2}$ $<$  $S_{\ell}^{2}$ are the so called evanescent zone with exponentially decaying behaviour. In Figures 4 and 5, we mark them with different grey shadings. Brunt$-$V$\ddot{\rm a}$is$\ddot{\rm a}$l$\ddot{\rm a}$ frequency $N$ is independent of $\ell$, while characteristic acoustic frequency $S_{\ell}$ is proportional to $\sqrt{\ell(\ell+1)}$. In Figures 4 and 5, it can be clearly seen that the propagation zones of g modes with $\ell$ = 2 are much thicker than those with $\ell$ = 1 in the observed frequency range. However, the evanescent zones between the propagation zones of p modes and the propagation zones of g modes for $\ell$ = 2 modes are much thinner than those for $\ell$ = 1 modes, which means that the propagation zones of p modes and the propagation zones of g modes for $\ell$ = 2 modes are poorer separated than those for $\ell$ = 1 modes.

The kinetic energy of one oscillation mode can be defined as
\begin{equation}
E_{\rm kin} = 2\pi\omega^{2}\int_{0}^{R}[\xi_{r}^{2}+\ell(\ell+1)\xi_{h}^{2}]\rho r^{2}dr,
\end{equation}
where the integrand is the kinetic energy weight function. Figure 6 illustrates profiles of the scaled kinetic energy weight function inside the best-fitting model. In Figures 4 and 5, it can be clearly seen that the propagation zones of g modes with $\ell$ = 1 and 2 located in the regions with 0.14 $<$ $r/R$ $<$ 0.28 and 0.14 $<$ $r/R$ $<$ 0.40, respectively. In Figure 6, it can be noticed that $\ell$ = 1 modes mainly propagate in the propagation zones of p modes, indicating that $\ell$ = 1 modes are p modes. However, those oscillation modes with $\ell$ = 2 behave propagating features of the mixed modes. They mainly propagate like p mode in the propagation zones of p modes. Meanwhile, they behave pronounced g-mode features in the propagation zones of g modes. This fact indicates that oscillation modes with $\ell$ = 2 are mixed modes, which are p-dominated modes with pronounced g-mode features. This might explain why the average frequency spacings of the two sets of multiplets are not equal.

\section{Summary and Conclusions}
In the present work, we try to determined precisely the evolutionary stage and physical parameters of HIP 80088 with the method of asteroseismology. We compute a grid of theoretical models to fit the twelve observed frequencies by performing a $\chi^{2}$ minimization. The main results are summarized as follows:

1. According to model calculations, we find that the equatorial rotation velocity $\upsilon_{\rm e}$ converges to 120 $-$ 136 km s$^{-1}$. The projected rotational velocity $\upsilon$sin$i$ is measured to be 80 $\pm$ 5 km s$^{-1}$( Ripepi et al 2015). Then the inclination angle $i$ of HIP 80088 is suggested to be 38.3 $_{-4.8}^{+6.8}$ degree.

2. The fundamental parameters of HIP 80088 are found to converge to a small range, i.e., $M$ = 1.68 $-$ 1.78 $M _{\odot}$, $Z$ = 0.015 $-$ 0.018, $\log g$ = 4.114 $-$ 4.125, $R$ = 1.882 $-$ 1.914 $R_{\odot}$, $\tau_{0}$ = 7636 $-$ 7723 s, and age = 9.03 $-$ 10.21 Myr. Meanwhile, we find that HIP 80088 is a pre-main sequence star, in which CN cycle has not arrived at the equilibrium state. The mass fractions of C12 and N14 in metal composition are determined to be 0.1277$^{+0.0064}_{-0.0049}$ and 0.1092$^{+0.0057}_{-0.0074}$ respectively. Twenty-eight percent of C12 have turned into N14.

3. Based on our model fittings, we suggest that $f_{3}$ is one radial mode, $f_{2}$, $f_{4}$, $f_{8}$, and $f_{11}$ are four $\ell$ = 1 modes, and $f_{1}$, $f_{5}$, $f_{6}$, $f_{7}$,$f_{9}$, $f_{10}$, and $f_{12}$ are seven $\ell$ = 2 modes. Moreover, we find that ($f_{2}$, $f_{4}$, $f_{8}$) form one complete triplet, and ($f_{5}$, $f_{7}$, $f_{9}$, $f_{10}$) form four components of one quintuplet.

4. Based on the best-fitting model, we find that $\ell$ = 2 modes of HIP 80088 are mixed modes, which are p-dominated modes with pronounced g-mode features. While for the oscillation modes with $\ell$ = 1, they are p modes. This may be the reason that two averaged frequency spacings in the two sets of multiplets are not equal.

\acknowledgments
This work is funded by the NSFC of China (Grant No. 11333006, 11521303, 11803082) and by the foundation of Light of West China Program from Chinese Academy of Sciences. The authors gratefully acknowledge the computing time granted by the Yunnan Observatories, and provided on the facilities at the Yunnan Observatories Supercomputing Platform. The authors are sincerely grateful to an anonymous referee for instructive advice and productive suggestions. The authors thank Jie Su, Qian-sheng Zhang, Tao Wu, and Wei-kai Zong for their fruitful suggestions.

\begin{figure*}
 \epsscale{1.0}
 \centering
 \plotone{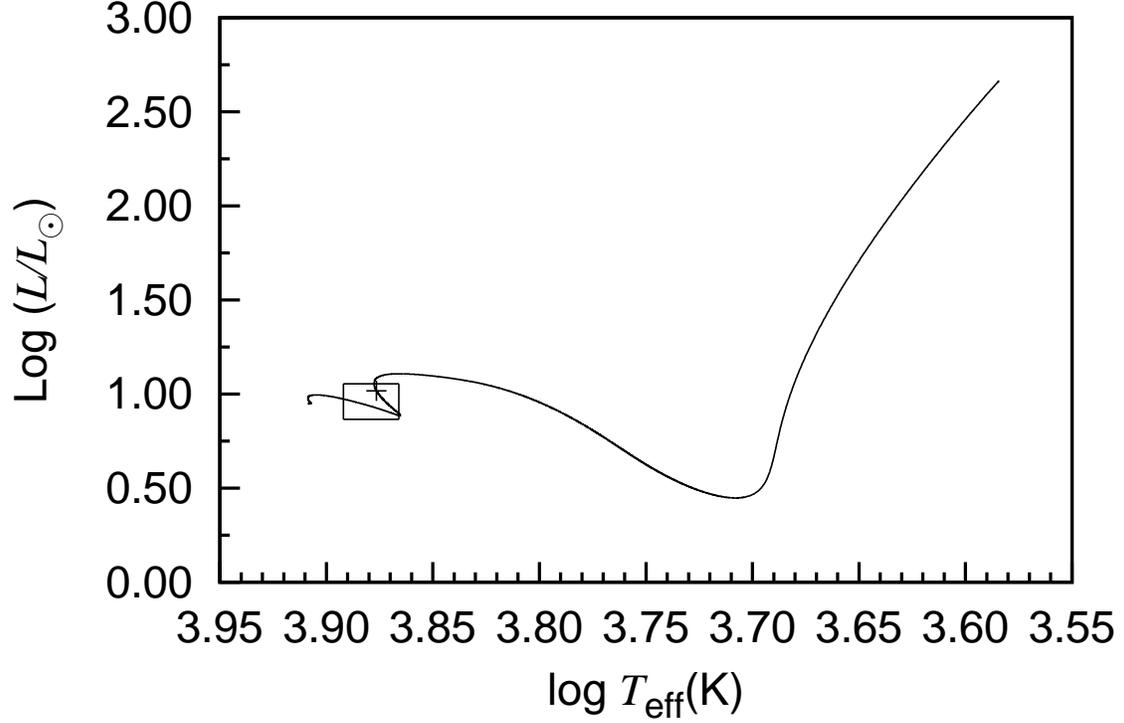}
  \caption{Evolutionary track of $M$ = 1.74 $M_{\odot}$ and  $Z$ = 0.017. The rectangle corresponds to the error box of the observed parameters, 0.865 $<$ $\log (L/L_{\odot})$ $<$ 1.055 and 7350 K $<$ $T_{\rm eff}$ $<$ 7800 K. The cross denotes the location of the best-fitting model (Model 29).}
  \label{Figure.1}
\end{figure*}

\begin{figure*}
  \epsscale{1.0}
  \centering
 \plotone{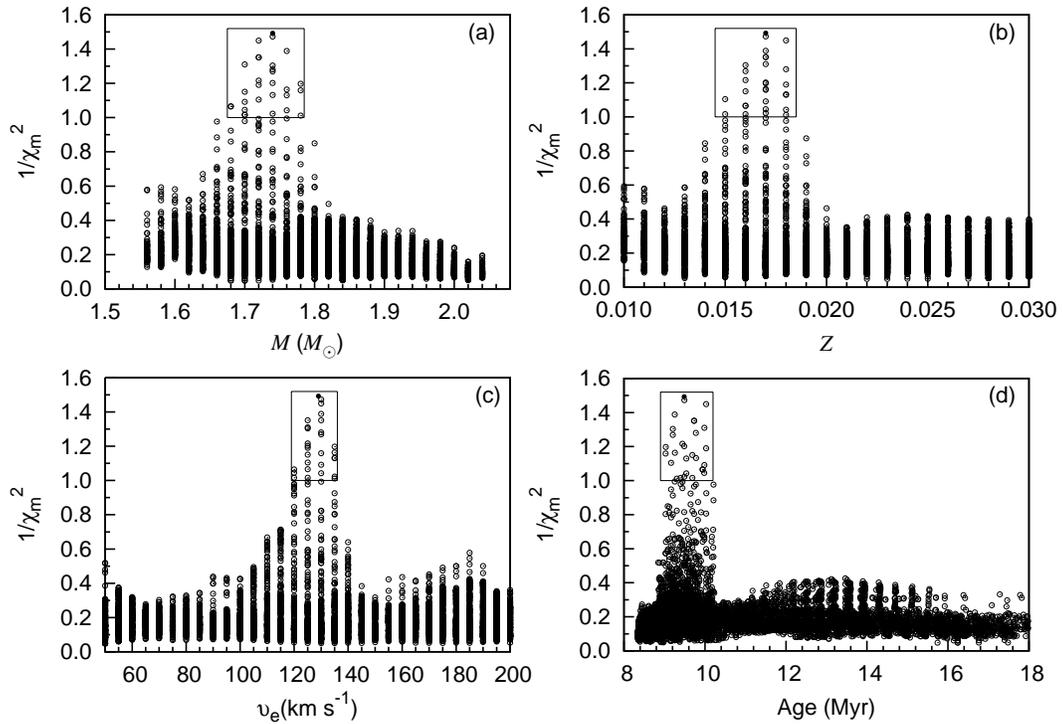}
  \caption{Visualisation of the resulting 1/$\chi_{\rm m}^{2}$ as a function of the stellar mass $M$ (Panel (a)), the metallicity $Z$ (Panel (b)), the equatorial rotation velocity $\upsilon_{\rm e}$ (Panel (c)), and the age (Panel (d)), respectively.}
  \label{Figure.2}
\end{figure*}

\begin{figure*}
  \epsscale{1.0}
   \centering
  \plotone{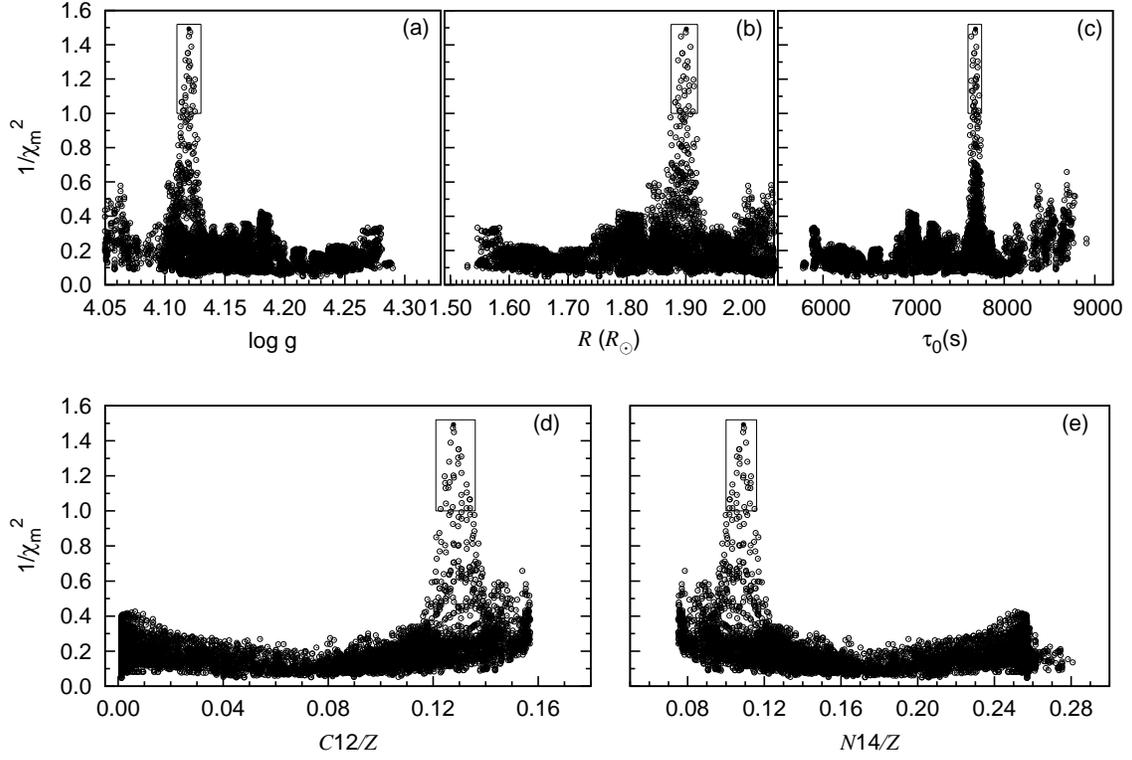}
  \caption{Visualisation of the resulting 1/$\chi_{\rm m}^{2}$ as a function of the gravitational acceleration $\log g$ (Panel (a)), the radius R (Panel (b)), the acoustic radius $\tau_{0}$ (Panel (c)), the mass fraction of C12 in metal composition C12/$Z$ (Panel (d)), and the mass fraction of N14 in metal composition N14/$Z$ (Panel (e)), respectively.}
  \label{Figure.3}
\end{figure*}

\begin{figure*}
  \epsscale{1.0}
   \centering
  \plotone{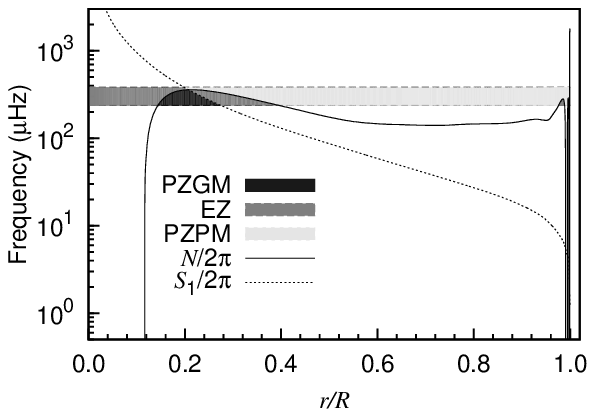}
  \caption{Brunt$-$V$\ddot{\rm a}$is$\ddot{\rm a}$l$\ddot{\rm a}$ frequency  $N$ and  characteristic acoustic frequencies $S_{\ell}$ ($\ell= 1$) inside the best-fitting model of HIP 80088. The two horizontal lines mark the frequency range of observed oscillation modes (between 237 and 382 $\mu$Hz ). PZGM represents the propagation zones of g modes, PZPM represents the propagation zones of p modes, and EZ represents the evanescent zone.}
  \label{Figure.4}
\end{figure*}

\begin{figure*}
  \epsscale{1.0}
   \centering
  \plotone{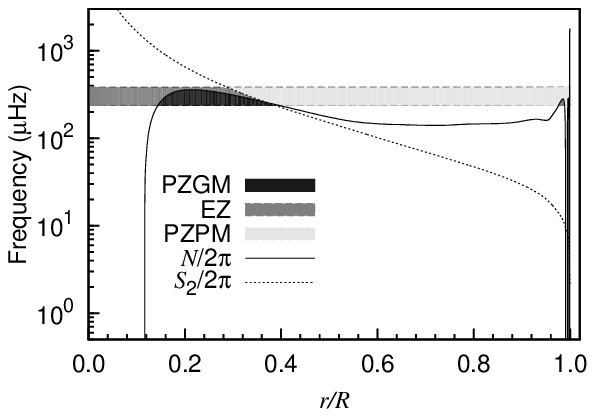}
  \caption{Brunt$-$V$\ddot{\rm a}$is$\ddot{\rm a}$l$\ddot{\rm a}$ frequency  $N$ and  characteristic acoustic frequencies $S_{\ell}$ ($\ell= 2$) inside the best-fitting model of HIP 80088. The two horizontal lines mark the frequency range of observed oscillation modes (between 237 and 382 $\mu$Hz ). PZGM represents the propagation zones of g modes, PZPM represents the propagation zones of p modes, and EZ represents the evanescent zone.}
  \label{Figure.5}
\end{figure*}

\begin{figure*}
  \epsscale{1.0}
     \centering
  \plotone{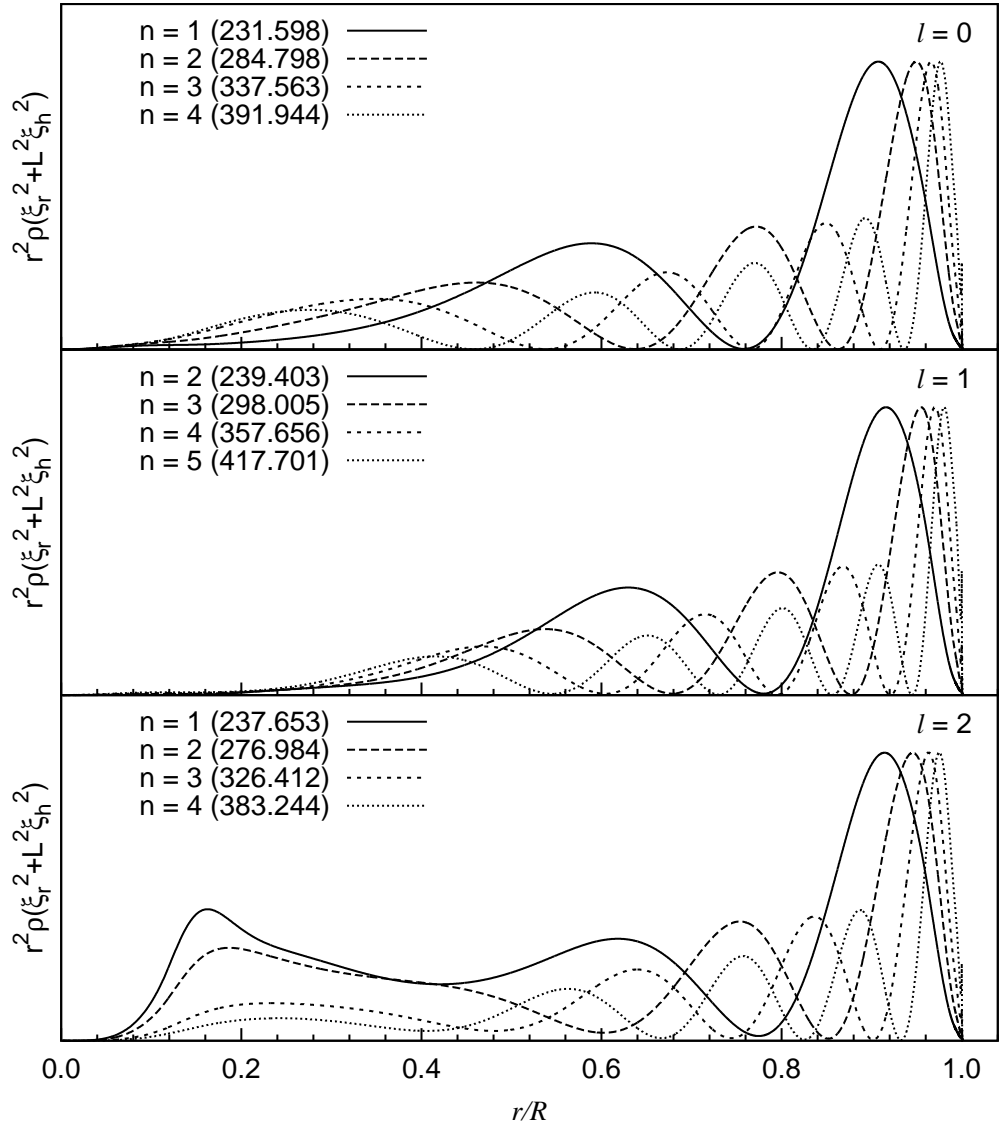}
  \caption{Scaled kinetic energy weight functions inside the best-fitting model.}
  \label{Figure.6}
\end{figure*}

\begin{table*}
\centering
\caption{\label{t1}Twelve independent frequencies of HIP 80088 obtained by Ripepi et al. (2015).}
\begin{tabular}{lccccc}
\hline\hline
ID            &Freq.          &Freq.               &Ampl.    \\
               &d$^{-1}$     &($\mu$Hz)     &(ppt)    \\
\hline
$f_{1}$   &20.5083     &237.365         &0.066    \\
$f_{2}$   &24.3411      &281.726         &0.156    \\
$f_{3}$   &24.5972     &284.690        &0.315    \\
$f_{4}$   &25.7598     &298.146        &0.869    \\
$f_{5}$   &25.9414     &300.248        &0.819    \\
$f_{6}$   &26.2301     &303.589        &0.098    \\
$f_{7}$   &26.9462     &311.877         &0.311    \\
$f_{8}$   &27.1136      &313.815         &1.073    \\
$f_{9}$   &28.2886     &327.414        &0.520    \\
$f_{10}$ &29.4115      &340.411         &0.227    \\
$f_{11}$  &31.0072     &358.880        &0.119    \\
$f_{12}$  &32.9888    &381.815         &0.770    \\
\hline
\end{tabular}
\end{table*}

\begin{table*}
%\footnotesize
\centering
\caption{\label{t2}Candidate models with $\chi_{\rm m}^{2}$ $<$ 1.0.}
\begin{tabular}{ccccccccccccccccc}
\hline\hline
Model &$\upsilon_{eq}$ &$Z$  &$M$   &$T_{\rm eff}$ &log($L/L_{\odot}$)   &$R$ &log$g$  &$\tau_{0}$  &$X_{\rm c}$ &$C12/Z$  &$N14/Z$ &Age &$\chi_{\rm m}^{2}$  \\
&(km s$^{-1}$)  &     &($M_{\odot}$) &(K) &  &$(R_{\odot})$  &(dex)  &$(\rm s)$ & & & &(Myr)  &  \\
\hline
1  &120 &0.016 &1.68 &7426 &0.985 &1.882 &4.114 &7674 &0.7142 &0.1339 &0.1019 & 9.94 &0.939\\
2  &120 &0.016 &1.70 &7494 &1.004 &1.889 &4.116 &7685 &0.7142 &0.1327 &0.1034 & 9.69 &0.984\\
3  &120 &0.017 &1.70 &7393 &0.980 &1.887 &4.117 &7659 &0.7116 &0.1309 &0.1055 & 9.98 &0.957\\
4  &125 &0.015 &1.70 &7595 &1.028 &1.891 &4.115 &7714 &0.7167 &0.1341 &0.1018 & 9.40 &0.984\\
5  &125 &0.015 &1.72 &7664 &1.047 &1.898 &4.117 &7723 &0.7167 &0.1328 &0.1032 & 9.16 &0.905\\
6  &125 &0.016 &1.68 &7426 &0.985 &1.882 &4.114 &7674 &0.7142 &0.1339 &0.1019 & 9.94 &0.938\\
7  &125 &0.016 &1.70 &7494 &1.004 &1.889 &4.116 &7685 &0.7142 &0.1327 &0.1034 & 9.69 &0.869\\
8  &125 &0.016 &1.72 &7562 &1.023 &1.896 &4.118 &7700 &0.7142 &0.1308 &0.1056 & 9.44 &0.822\\
9  &125 &0.016 &1.74 &7628 &1.041 &1.903 &4.120 &7706 &0.7142 &0.1295 &0.1071 & 9.21 &0.767\\
10 &125 &0.017 &1.70 &7393 &0.979 &1.887 &4.117 &7654 &0.7116 &0.1307 &0.1057 & 9.98 &0.763\\
11 &125 &0.017 &1.72 &7458 &0.998 &1.894 &4.119 &7668 &0.7116 &0.1297 &0.1069 & 9.73 &0.740\\
12 &125 &0.017 &1.74 &7523 &1.016 &1.901 &4.120 &7682 &0.7116 &0.1277 &0.1092 & 9.49 &0.832\\
13 &125 &0.018 &1.72 &7354 &0.973 &1.892 &4.120 &7636 &0.7091 &0.1279 &0.1089 &10.03 &0.840\\
14 &130 &0.016 &1.74 &7628 &1.041 &1.903 &4.120 &7706 &0.7142 &0.1295 &0.1071 & 9.21 &0.788\\
15 &130 &0.017 &1.70 &7393 &0.979 &1.887 &4.117 &7654 &0.7116 &0.1307 &0.1057 & 9.98 &0.916\\
16 &130 &0.017 &1.72 &7458 &0.998 &1.894 &4.119 &7668 &0.7116 &0.1297 &0.1069 & 9.73 &0.740\\
17 &130 &0.017 &1.74 &7523 &1.016 &1.900 &4.121 &7677 &0.7116 &0.1276 &0.1094 & 9.49 &0.679\\
18 &130 &0.017 &1.76 &7587 &1.034 &1.908 &4.122 &7691 &0.7116 &0.1266 &0.1104 & 9.25 &0.720\\
19 &130 &0.017 &1.78 &7650 &1.051 &1.914 &4.124 &7695 &0.7116 &0.1246 &0.1129 & 9.03 &0.863\\
20 &130 &0.018 &1.72 &7354 &0.973 &1.892 &4.120 &7636 &0.7091 &0.1279 &0.1089 &10.03 &0.690\\
21 &130 &0.018 &1.74 &7417 &0.991 &1.899 &4.121 &7650 &0.7091 &0.1261 &0.1111 & 9.78 &0.781\\
22 &130 &0.018 &1.76 &7479 &1.008 &1.905 &4.123 &7657 &0.7091 &0.1249 &0.1125 & 9.54 &0.959\\
23 &135 &0.017 &1.74 &7522 &1.015 &1.900 &4.121 &7672 &0.7116 &0.1274 &0.1095 & 9.49 &0.978\\
24 &135 &0.017 &1.76 &7586 &1.033 &1.906 &4.123 &7681 &0.7116 &0.1263 &0.1108 & 9.26 &0.859\\
25 &135 &0.017 &1.78 &7650 &1.051 &1.913 &4.125 &7690 &0.7116 &0.1244 &0.1131 & 9.03 &0.835\\
26 &135 &0.018 &1.74 &7416 &0.990 &1.898 &4.122 &7641 &0.7091 &0.1257 &0.1115 & 9.78 &0.883\\
27 &135 &0.018 &1.76 &7478 &1.008 &1.905 &4.124 &7652 &0.7091 &0.1247 &0.1127 & 9.54 &0.885\\
28 &135 &0.018 &1.78 &7540 &1.025 &1.912 &4.125 &7665 &0.7091 &0.1228 &0.1149 & 9.31 &0.989\\
\textbf{29} &\textbf{129} &\textbf{0.017} &\textbf{1.74} &\textbf{7523} &\textbf{1.016} &\textbf{1.901} &\textbf{4.120} &\textbf{7682} &\textbf{0.7116} &\textbf{0.1277} &\textbf{0.1092} &\textbf{9.49} &\textbf{0.670}\\
\hline
\end{tabular}
\end{table*}

\begin{table*}
\centering
\caption{\label{t3} Fundamental parameters of HIP 80088.}
\begin{tabular}{lll}
\hline\hline
Parameter               &Values \\
\hline
$M(M_{\odot})$            &1.74$^{+0.04}_{-0.06}$\\
$Z$                               &0.017 $^{+0.001}_{-0.002}$\\
$T_{\rm eff}$ (K)          &7523  $\pm$ 170 \\
log$g$ (dex)                &4.120 $^{+0.005}_{-0.006}$\\
$R(R_{\odot})$             &1.901 $^{+0.013}_{-0.019}$\\
log($L/L_{\odot}$)       &1.016$^{+0.031}_{-0.043}$\\
Age (Myr)                     &9.49$^{+0.72}_{-0.46}$\\
$\upsilon_{\rm eq}$(km s$^{-1}$)    &129 $^{+7}_{-9}$  \\
$\tau_{0}$(s)               &7682$^{+41}_{-46}$\\
$C12/Z$                      &0.1277$^{+0.0064}_{-0.0049}$\\
$N14/Z$                      &0.1092$^{+0.0057}_{-0.0074}$  \\
\hline
\end{tabular}
\end{table*}

\begin{table*}
\centering
\caption{\label{t4}Theoretical frequencies of the best-fitting model.}
\label{observed frequencies}
\begin{tabular}{llllll}
\hline\hline
$\nu^{\rm theo}(\ell,n)$ &$\beta_{\ell, n}$ &$\nu^{\rm theo}(\ell,n)$ &$\beta_{\ell, n}$ &$\nu^{\rm theo}(\ell,n)$ &$\beta_{\ell, n}$\\
\hline
179.414(0,0) &  & 183.899(1,1) & 0.988 & 197.897(2,0) & 0.930 \\
231.598(0,1) &  & 239.403(1,2) & 0.996 & 237.653(2,1) & 0.973 \\
284.798(0,2) &  & 298.005(1,3) & 0.996 & 276.984(2,2) & 0.849 \\
337.563(0,3) &  & 357.656(1,4) & 0.993 & 326.412(2,3) & 0.887 \\
391.944(0,4) &  & 417.701(1,5) & 0.990 & 383.244(2,4) & 0.932 \\
448.647(0,5) &  & 476.770(1,6) & 0.988 & 442.369(2,5) & 0.955 \\
\hline
\end{tabular}
\end{table*}

\begin{table*}
\centering
\caption{\label{t5}Comparing results between theoretical frequencies of the best-fitting model and the observations.}
\begin{tabular}{ccclc}
\hline\hline
ID  &$\nu^{\rm obs}$   &$\nu^{\rm theo}$ &($\ell$, $n$, $m$) & $|\nu^{\rm obs}-\nu^{\rm theo}|$\\
     &($\mu$Hz)         &($\mu$Hz)              &              &($\mu$Hz)\\
\hline
$f_{1}$    &237.365           &237.653          &(2, 1, 0)          &0.288\\
$f_{2}$    &281.726           &282.552         &(1, 3, -1)          &0.826\\
$f_{3}$    &284.690          &284.798          &(0, 2, 0)          &0.108\\
$f_{4}$    &298.146           &298.005          &(1, 3, 0)          &0.141\\
$f_{5}$    &300.248          &298.888          &(2, 3, -2)        &1.360\\
$f_{6}$    &303.589          &303.329          &(2, 2, +2)       &0.260\\
$f_{7}$    &311.877           &312.650           &(2, 3, -1)          &0.773\\
$f_{8}$    &313.815           &313.458          &(1, 3, +1)         &0.357\\
$f_{9}$    &327.414           &326.412          &(2, 3, 0)          &1.002\\
$f_{10}$   &340.411           &340.174          &(2, 3, +1)         &0.237\\
$f_{11}$   &358.880           &357.656         &(1, 4, 0)           &1.224\\
$f_{12}$   &381.815           &383.244          &(2, 4, 0)          &1.429\\
\hline
\end{tabular}
\end{table*}

\begin{table*}
\centering
\caption{\label{t6} Possible rotational splitting.}
\begin{tabular}{lccllccc}
\hline\hline
Mult.&ID &$\nu^{\rm obs}$ &$\delta\nu$  &$\nu^{\rm theo}$($\beta_{\ell,n}$)  &($\ell$, $n$, $m$) & $|\nu^{\rm obs}-\nu^{\rm theo}|$\\
     & &($\mu$Hz)         &($\mu$Hz)    &($\mu$Hz)      & &($\mu$Hz)                   \\
\hline
      &$f_{2}$  &281.726   &               &282.552               &(1, 3, -1)         &0.826\\
      &              &                 &16.420   &\\
1    &$f_{4}$  &298.146   &               &298.005(0.996)   &(1, 3, 0)          &0.141\\
     &               &                  &15.669   &\\
      &$f_{8}$ &313.815   &               &313.458              &(1, 3, +1)         &0.357\\
     &&\\
     &&\\
     &$f_{5}$  &300.248   &               &298.888              &(2, 3, -2)         &1.360\\
     &              &                 &11.629    &\\
     &$f_{7}$  &311.877    &               &312.650               &(2, 3, -1)         &0.773\\
     &              &                 &15.537    &\\
2   &$f_{9}$  &327.414   &               &326.412(0.887)  &(2, 3, 0)          &1.002\\
     &               &                 &12.997   &\\
     &$f_{10}$   &340.411    &              &340.174               &(2, 3, +1)         &0.237\\
\hline
\end{tabular}
\end{table*}

\end{document}